\documentclass[12pt,manuscript]{elsart}
\usepackage{epsfig,harvard,amssymb,rotating,lscape}
\newcommand{\DD}{\displaystyle}
\newcommand{\D}{\displaystyle}

\makeatletter
\def\elsartstyle{%
        \def\normalsize{\@setfontsize\normalsize\@xiipt{14.5}}
        \def\small{\@setfontsize\small\@xipt{13.6}}
        \let\footnotesize=\small
        \def\large{\@setfontsize\large\@xivpt{18}}
        \def\Large{\@setfontsize\Large\@xviipt{22}}
        \skip\@mpfootins = 18\p@ \@plus 2\p@
        \normalsize
}
\makeatother

\def\ar{{Annu. Rev. Astron. Astrophys.} \,}

\def\apj{{Ap. J.} \,}

\def\apjl{{Ap. J. Lett.} \,}

\def\mn{{MNRAS} \,}
\def\na{{New Astronomy} \,}

\def\ea{et al. \,}
\def\bibcode#1{(\texttt{#1})} 
\def\eg{{e.g.\ }}

\begin{document}
                       
\begin{frontmatter}
\title{S-Z Anisotropy \& Cluster Counts: Consistent \\ 
Selection of $\sigma_8$ \& the Temperature-Mass Relation}

\author{Sharon Sadeh\thanksref{email}}
\address{School of Physics and Astronomy, Rayomnd and Beverly Sackler
Faculty of Exact Sciences, Tel Aviv University, Tel Aviv, 69978, Israel}

\thanks[email]{E-mail: shrs@post.tau.ac.il}

\and

\author{Yoel Rephaeli}
\address{School of Physics and Astronomy, Raymond and Beverly Sackler
Faculty of Exact Sciences, Tel Aviv University, Tel
Aviv, 69978, Israel, \\and\\ Center for Astrophysics and Space
Sciences, University of California, San Diego, La Jolla,
CA\,92093-0424}

\begin{abstract}       

The strong dependence of the mass variance parameter, $\sigma_8$, on 
the adopted cluster mass-temperature relation is explored. A recently 
compiled X-ray cluster catalog and various mass-temperature relations 
are used to derive the corresponding values of $\sigma_8$. Calculations
of the power spectrum of the CMB anisotropy induced by the 
Sunyaev-Zeldovich effect and cluster number counts are carried out 
in order to assess the need for a consistent choice of the 
mass-temperature scaling and the parameter $\sigma_8$. We find that 
the consequences of inconsistent choice of the mass-temperature 
relation and $\sigma_8$ could be quite substantial, including a 
considerable mis-estimation of the magnitude of the power spectrum and 
cluster number counts. Our results can partly explain the large scatter 
between published estimates of the power spectrum and number counts.
We also show that the range of values of the power-law index in the 
scaling $C_{\ell}\sim\sigma_8^{6-7}$ deduced in previous studies is 
likely overestimated; we obtain a more moderate dependence,
$C_{\ell}\sim\sigma_8^{4}$. 
\end{abstract}

\end{frontmatter}

\section{Introduction}
\label{sect:intro}

Calculations of the power spectrum of CMB anisotropy induced by  
the Sunyaev-Zeldovich (S-Z) effect, and of cluster number counts, 
involve various intrinsic cluster quantities, as well as cosmological 
and cluster mass function parameters. Specifically, the level of 
S-Z signal in a cluster is directly related to the intracluster (IC) 
gas temperature and density distributions. The cosmological and large 
scale structure parameters enter in the expression for the collective 
mass function of clusters. Most commonly, this function is assumed to 
have the Press \& Schechter (1974) form, but other, more physically 
established mass functions, are also widely adopted. The mass 
function is usually normalized by specifying an observationally 
deduced value of $\sigma_8$, the rms density fluctuation on a scale 
of $8\,h^{-1}$ Mpc. The value of $\sigma_8$ may be deduced in 
several different ways, perhaps most directly from a comparison 
between the observed two-point correlation function of galaxies or 
clusters of galaxies, and the spectrum of the primordial density 
fluctuation field on the relevant scales, recently employed in 
the analysis of the large SDSS sample (\eg, Tegmark \ea 2003).
Measurements of CMB anisotropy yield a globally averaged value 
of $\sigma_8$, most recently deduced from the first year WMAP 
measurements (Bennett et al., 2003). Additionally, $\sigma_8$ has 
been deduced from fitting X-ray cluster data (luminosity or 
temperature functions) to the corresponding theoretical models, 
with the latter derived from a mass function upon assuming a 
specific mass-luminosity or mass-temperature relation. 

Deduced values of $\sigma_8$ clearly depend on data quality, 
modeling, and analysis methods, as is reflected in the current 
wide range, $0.5\lesssim\sigma_8\lesssim 1.2$. Joint analysis of 
the SDSS and WMAP databases yields the current `best' (with 6 
free `vanilla' parameters) value, $\sigma_8=0.917^{+0.090}_{-0.072}$
(Tegmark et al. 2003). This is roughly in agreement (e.g., Wu 2001)
or somewhat lower (Viana \& Liddle 1999) than results from X-ray
cluster studies. The wide range of values of $\sigma_8$ has
significant implications on the level of various cosmological
quantities. Here we explore some of its consequences for the
abundance of clusters, and in particular the power of S-Z
anisotropy and cluster number counts, quantities that depend
steeply on $\sigma_8$. Since we concentrate here on clusters,
the emphasis will be on values of $\sigma_8$ deduced from cluster
X-ray measurements, particularly of gas temperatures. Pertinent
key features of such analyses are properties of the cluster sample 
(\eg, completeness, temperature and redshift ranges), the adopted 
mass function model, and the mass-temperature relation used to 
convert the mass function into a temperature function. By 
comparing the latter function to the observed temperature function, 
$\sigma_8$ (and other parameters, such as $\Omega_m$) can 
be determined.

In this paper we focus on consequences of the choice of the 
temperature-mass (TM) relation on $\sigma_8$, and 
their impact on the predicted S-Z angular power spectrum and 
cluster number counts. A quantitative assessment is given of 
the impact of inconsistent choice of the TM relation and 
$\sigma_8$ on the predicted levels of the the S-Z anisotropy 
and the number of clusters to be observed by the Planck 
satellite. Our main goal here is to improve some of the 
technical aspects in the calculations of these quantities, 
and to partly explain the large variance among previous 
estimates of the S-Z power spectrum.

The paper is arranged as follows: the need for a consistent 
choice of $\sigma_8$ and the TM relation, the method employed 
to extract $\sigma_8$ from the observed temperature function,
and the variation of $\sigma_8$ with the choice of the TM relation
are described in \S\ 2. The importance of consistent choice is 
addressed once more in \S\ 3 and \S\ 4, where its consequences on 
the S-Z angular power spectrum and cluster number counts are
quantified and further elaborated upon in \S\ 5.

\section{TM-$\sigma_8$ Relation}
\label{sect:sig-mt}

\subsection{Method}

The calculation of the S-Z power spectrum involves the basic 
cosmological and large structure parameters, properties of IC gas, 
and the cluster mass function (MF). Of the various methods that 
have been used to calculate the S-Z anisotropy, we employ the 
approach described in Colafrancesco \ea (1994, 1997), which 
we have recently updated and generalized (Sadeh \& Rephaeli 2003). 
The calculation will not be described here; full details are given 
in the latter paper.

In a cluster temperature-mass relation (TMR) the gas temperature 
is a monotonically increasing function of mass; $T\propto M^{2/3}$ 
in virial equilibrium. The temperature function may be written in 
terms of the MF by 
\begin{equation}
N(T)\,dT=N[M(T)]\,\frac{dT}{dM}\,dM.
\end{equation}
Since at the high mass end the MF is to a good approximation 
a monotonically increasing function of $\sigma_8$, it can be 
formally shown (Sadeh \& Rephaeli 2003) that:

\noindent
1. TM relations which predict higher (lower) temperatures for a given
mass yield lower (higher) $\sigma_8$, and therefore, lower (higher)
cluster density.

\noindent
2. Calculations of the S-Z power spectrum employing a specific TMR,  
and a value of $\sigma_8$ different from the one that would have 
been extracted by adopting this relation, will result in 
mis-estimation of S-Z power. The power level will be overestimated 
(underestimated) when, for a given TMR, the adopted value of 
$\sigma_8$ is higher (lower) than the value corresponding to 
that particular TMR.

We adopt the procedure developed by Pierpaoli et al. (2001) for 
the determination of $\sigma_8$ from an observationally based 
temperature function. Given a list of X-ray cluster temperatures 
and their corresponding redshifts, the temperature function can 
be deduced. The theoretical MF -- with $\sigma_8$ as the free 
parameter -- is converted into a temperature function given an 
assumed TM relation. This is calculated at redshift $z=0.057$, 
the median redshift of the particular X-ray catalog used. The 
temperature range in the catalog is divided into a large number 
of bins, and the mean occupation number of clusters in each of 
these bins is computed by multiplying $n(T)\,dT$ at the bin 
temperature by the effective volume occupied at this temperature, 
i.e., the volume of space to which clusters in the bin can be 
observed, denoted by $V_{i}$ . A large number of mock catalogs is 
now devised by using the listed temperatures as the mean of a 
normally distributed variable, and their $1\sigma$ errors as its 
variance. The temperature bins are now scanned so as to determine 
whether any of these is occupied by clusters in the mock catalog. 
Occupation numbers are assigned to each one of the bins. Assuming 
a Poissonian distribution of the cluster occupation numbers, the 
probability of finding $N_i$ clusters in bin $i$ is estimated 
using the Poissonian distribution, with $n(T)\,dT\,V_{i}$ interpreted 
as the distribution mean. Finally, the likelihood function of 
obtaining the `observed' occupations in all bins simultaneously is 
constructed, and maximized against $\sigma_8$.

\subsection{TM Relations}

We employ five TM relations found in the literature; three of these 
have the same functional form, differing only in the temperature 
normalization for a fiducial cluster at redshift $z=0$ with mass 
$10^{15}\,h^{-1}\,M_{\odot}$. The five functions are identified here 
as models 1-5:

\emph{Model 1 (Colafrancesco et al 1997):}

\begin{equation}
T=5.78\,(1+z)\,\left(\frac{M}{10^{15}\,h^{-1}\,M_{\odot}}\right)^{2/3}
\Omega_0^{1/3}\left[\frac{\Delta(\Omega_0,z)}{\Delta(\Omega_0=1,z=0)}\right]^{1/3}.
\end{equation}

\emph{Model 2 (Tomita 2003):}

\begin{equation}
T=5.2\,\gamma\,(1+z)\,\left(\frac{M}{10^{15}\,h^{-1}\,M_{\odot}}\right)^{2/3}
\Omega_0^{1/3}\left[\frac{\rho_{vir}(z)}{18\pi^{2}}\right]^{1/3},
\end{equation}

where $\gamma=1.2$.

\emph{Model 3 (Henry 2000):}

\begin{equation}
T=\frac{7.98\,keV}{\beta_{TM}}\,(1+z)\,\left(\frac{M}{10^{15}\,h^{-1}\,M_{\odot}}\right)^{2/3}
\Omega_0^{1/3}\left[\frac{\Delta(\Omega_0,z)}{18\pi^{2}}\right]^{1/3},
\end{equation}

where $\beta_{TM}=1.21$.

\emph{Model 4 (Pierpaoli et al. 2001):}

\begin{equation}
T=\beta\,\left(\frac{M}{10^{15}\,h^{-1}\,M_{\odot}}\right)^{2/3}
(\Delta_c\,E^{2})^{1/3}\,\left(1-2\frac{\Omega_{\Lambda}(z)}{\Delta_c}\right),
\end{equation}

where $E^{2}=\Omega_m(1+z)^{3}+\Omega_{\Lambda}$.

\emph{Model 5 (Molnar \& Birkinshaw 2000):}

\begin{equation}
T=2.76\,\beta^{-1}\frac{1-\Omega_0}{\Omega_0^{2/3}}\left(\frac{M}{10^{15}\,h^{-1}\,M_{\odot}}
\right)^{2/3}\left[\left(\frac{2\pi}{\sinh{\eta}-\eta}\right)^{2/3}+\frac{n_p+3}{5}\right],
\end{equation}

where $\beta$ and $n_{p}$ are taken to be $1$ and $-1.4$, respectively,
($n_p$ is the `effective' power index on cluster scales),
$\eta=\cosh^{-1}\left[\DD\frac{2}{\Omega_m(z)-1}\right]$, and 
$\Omega_m(z)=\DD\frac{\Omega_0(1+z)}{\Omega_0(1+z)+(1-\Omega_0)}$.

The first three relations may all be written in the form 
\begin{equation}
T=\alpha\,(1+z)\,\left(\frac{M}{10^{15}\,h^{-1}\,M_{\odot}}\right)^{2/3}
\,\Omega_0^{1/3}\,\left[\frac{\Delta_c(\Omega_0,z)}{18\pi^{2}}\right]^{1/3},
\label{eq:commonmt}
\end{equation}
where $\alpha=5.78, 6.24,6.59$ in models (1),(2), (3), respectively. 
In a flat universe with $\Lambda =0$, $\alpha$ is equal to the 
temperature of a local cluster with mass $10^{15}\,h^{-1}\,M_{\odot}$. 
In the currently fashionable (`standard') $\Lambda CDM$ model, 
substituting $z=0.053$ in the above relations, the temperatures 
associated with clusters having this mass are $3.43\,keV, 3.71
\,keV, 3.92\,keV,6.19\,keV,8.49\,keV$, respectively.

\subsection{Results for $\sigma_8$}

\begin{figure}
\centering
\epsfig{file=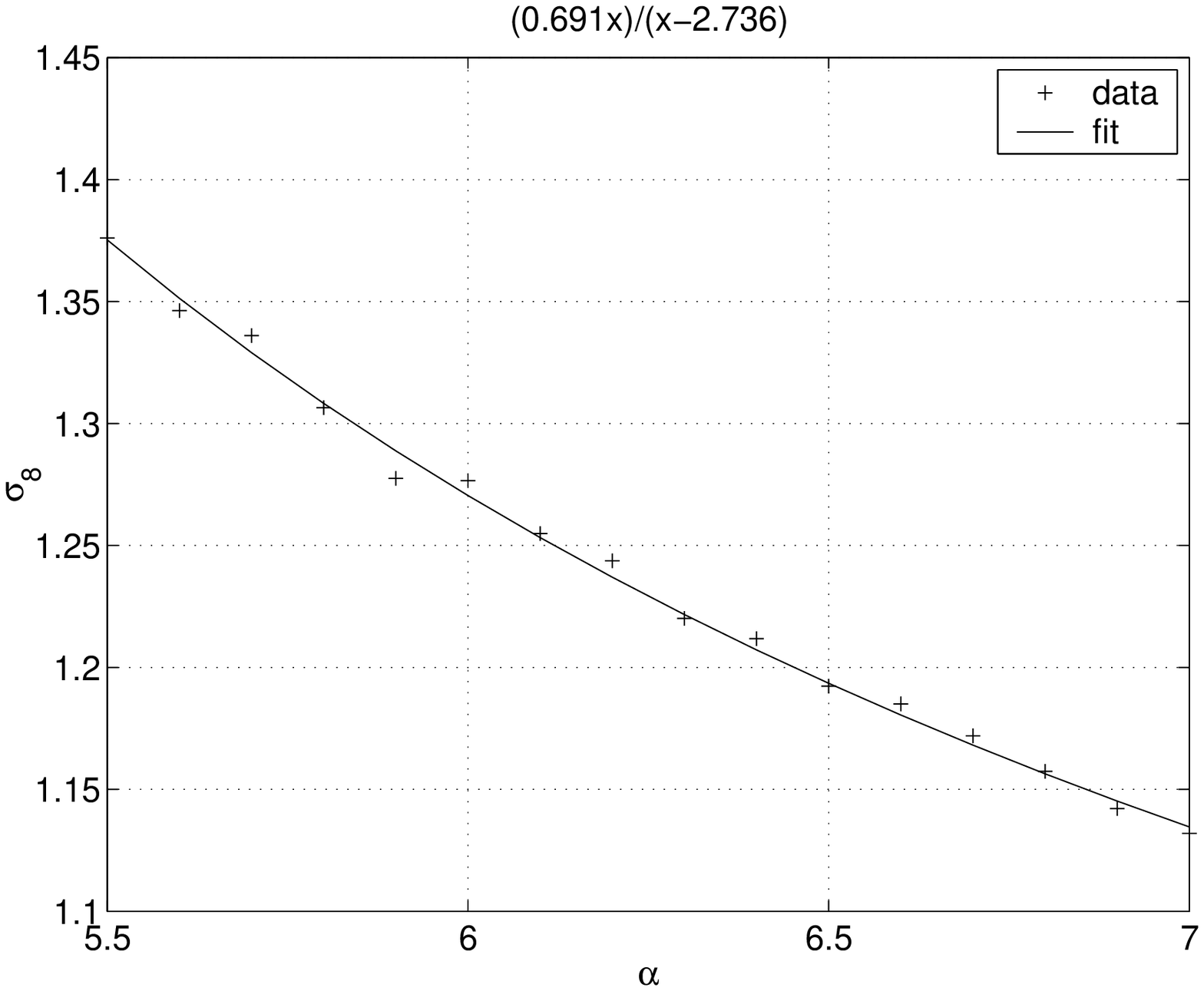, height=10cm, width=14cm, clip=}
\caption{$\sigma_8$ as function of the parameter $\alpha$. Each data
  point represents an average over ten runs. The fitted curve is
  plotted only for visual purposes.}
\label{fig:t-sig}
\end{figure}

To verify the validity of the first assertion made in the previous 
subsection, $\sigma_8$ has been estimated using equation~\ref{eq:commonmt} 
with $5.5\le\alpha\le 7.0$. For each value of $\alpha$ ten runs were made 
and average values of $\sigma_8$ were computed. A more statistically
complete analysis would require a much larger number of runs for each
value of $\alpha$, but the consistency of the results suggests that this
is not needed for our purposes here. The input and output of the runs are
tabulated in table 1 and illustrated in figure~\ref{fig:t-sig}. It is
clear from the plot that (as expected) $\sigma_8$ is quite sensitive
to the choice of the TM relation. In fact, a relative variance of
$\sim 30\%$ in the temperature scaling -- which is in accord with
typical observational uncertainties -- induces a relative change of
$\sim 20\%$ in the value of $\sigma_8$. This has important practical
consequences given that the mass function is extremely sensitive to
this parameter. Moreover, these results are in accordance with the
qualitative argument summarized above: A TM relation which associates
a higher temperature with a given mass yields a lower value of
$\sigma_8$, and vice versa. The calculated five values of
$\sigma_8$ are $1.32, \, 1.23, \, 1.18, \,1.00, \, 0.68$, for
models 1-5, respectively.

\begin{sidewaystable}[p]
\begin{tabular}{|c||c|c|c|c|c|c|c|c|c|c|c|c|c|c|c|c|}
\hline
{T (keV)}&{5.5}&{5.6}&{5.7}&{5.8}&{5.9}&{6.0}&{6.1}&{6.2}&{6.3}&{6.4}&{6.5}&{6.6}&{6.7}&{6.8}&{6.9}&{7.0} \\
\hline
\hline
{1}&{1.366}&{1.335}&{1.325}&{1.325}&{1.263}&{1.284}&{1.253}&{1.222}&{1.212}&{1.232}&{1.181}&{1.181}&{1.181}&{1.150}&{1.161}&{1.150} \\
{2}&{1.386}&{1.345}&{1.325}&{1.294}&{1.284}&{1.294}&{1.273}&{1.222}&{1.232}&{1.222}&{1.161}&{1.191}&{1.181}&{1.150}&{1.150}&{1.130} \\
{3}&{1.386}&{1.366}&{1.345}&{1.325}&{1.284}&{1.284}&{1.243}&{1.253}&{1.222}&{1.232}&{1.202}&{1.181}&{1.171}&{1.140}&{1.140}&{1.130} \\
{4}&{1.397}&{1.345}&{1.366}&{1.325}&{1.294}&{1.294}&{1.253}&{1.253}&{1.222}&{1.222}&{1.202}&{1.202}&{1.181}&{1.171}&{1.140}&{1.120} \\
{5}&{1.386}&{1.325}&{1.345}&{1.285}&{1.294}&{1.263}&{1.273}&{1.243}&{1.222}&{1.212}&{1.181}&{1.181}&{1.181}&{1.140}&{1.130}&{1.130} \\
{6}&{1.345}&{1.356}&{1.294}&{1.315}&{1.263}&{1.253}&{1.232}&{1.232}&{1.212}&{1.202}&{1.181}&{1.181}&{1.161}&{1.190}&{1.120}&{1.130} \\
{7}&{1.376}&{1.366}&{1.335}&{1.294}&{1.273}&{1.284}&{1.243}&{1.253}&{1.212}&{1.202}&{1.212}&{1.181}&{1.161}&{1.161}&{1.161}&{1.140} \\
{8}&{1.356}&{1.314}&{1.325}&{1.315}&{1.304}&{1.273}&{1.253}&{1.243}&{1.212}&{1.222}&{1.191}&{1.191}&{1.191}&{1.150}&{1.150}&{1.140} \\
{9}&{1.366}&{1.366}&{1.356}&{1.284}&{1.263}&{1.294}&{1.253}&{1.263}&{1.212}&{1.202}&{1.191}&{1.181}&{1.161}&{1.161}&{1.150}&{1.130} \\
{10}&{1.397}&{1.345}&{1.345}&{1.304}&{1.253}&{1.243}&{1.273}&{1.253}&{1.243}&{1.170}&{1.222}&{1.181}&{1.151}&{1.161}&{1.120}&{1.120} \\
\hline
{$<\sigma_8>$}&{1.376}&{1.346}&{1.336}&{1.307}&{1.278}&{1.277}&{1.255}&{1.244}&{1.220}&{1.212}&{1.192}&{1.185}&{1.172}&{1.157}&{1.142}&{1.132} \\
\hline
\end{tabular}
\label{table:sig8-t}
\caption{Values of $<\sigma_8>$ for the adopted TM relation}
\end{sidewaystable}

\section{Dependence of the Power Spectrum on $\sigma_8$ and the TMR}
\label{sect:cl}

\begin{figure}
\centering
\epsfig{file=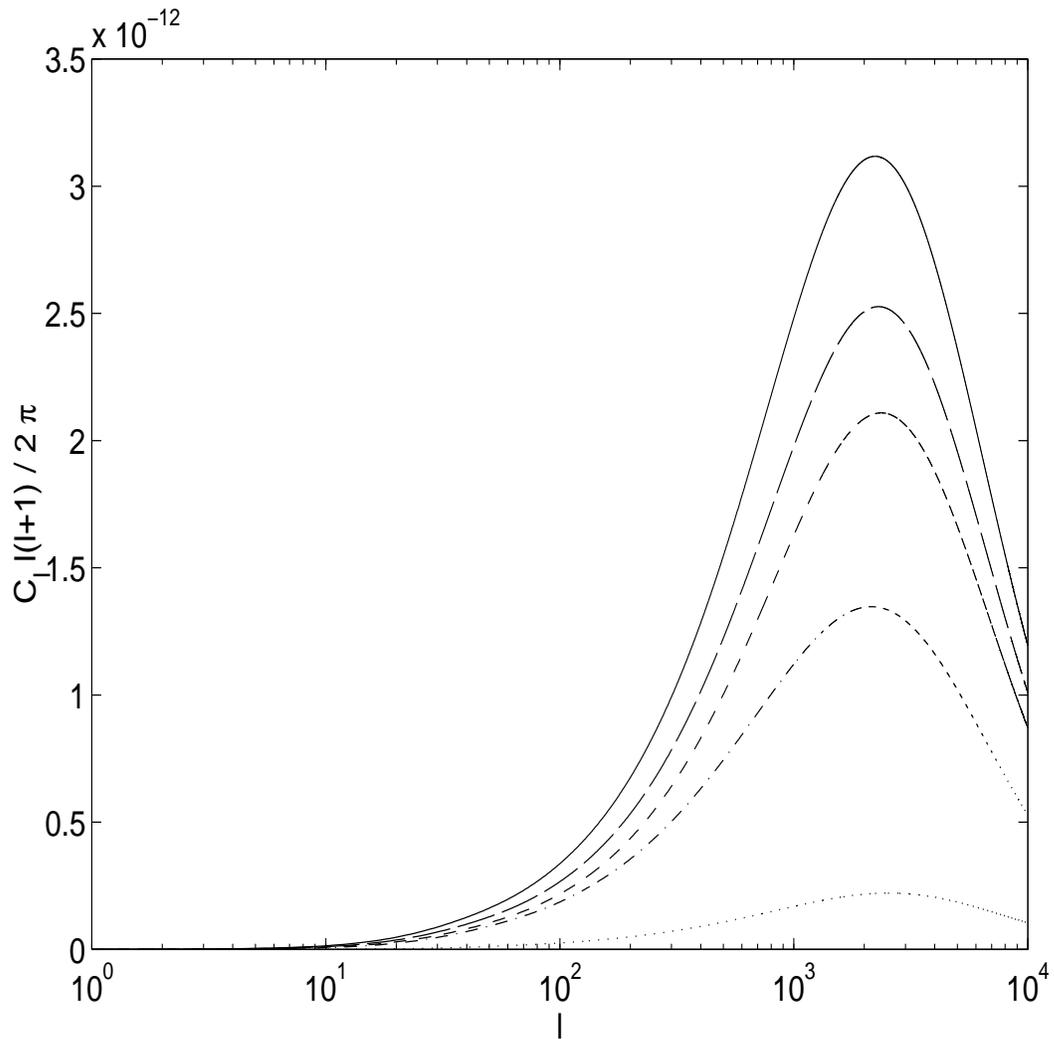, height=14cm, width=14cm, clip=}
\caption{The angular power spectrum of the S-Z effect calculated using 
the five M-T relations of models 1-5, and their matching
$\sigma_8$s as presented in \S\ 3. Solid, long dashed, short dashed,
dashed-dotted and dotted curves correspond to models 1,2,3,4 and 5,
respectively.}
\label{fig:cl}
\end{figure}

The S-Z power spectrum was calculated (Sadeh \& Rephaeli 2003) employing
the five models described above. We first calculated the power for each
of the five TM relations and its matching $\sigma_8$, and then crossed
these inputs by taking the TM relation of model $i$ and $\sigma_8$
corresponding to model $j$, where $i\ne j =1-5$. Results of the first
calculations are shown in figure~\ref{fig:cl}; these reflect the tight
relation between the magnitude of the power spectrum and the cluster
population. Note that although lower temperatures (implied by a `colder'
TM calibration) will tend to reduce the power, the influence of the
higher $\sigma_8$ dominates, resulting in a higher level of power.

\begin{figure}
\centering
\epsfig{file=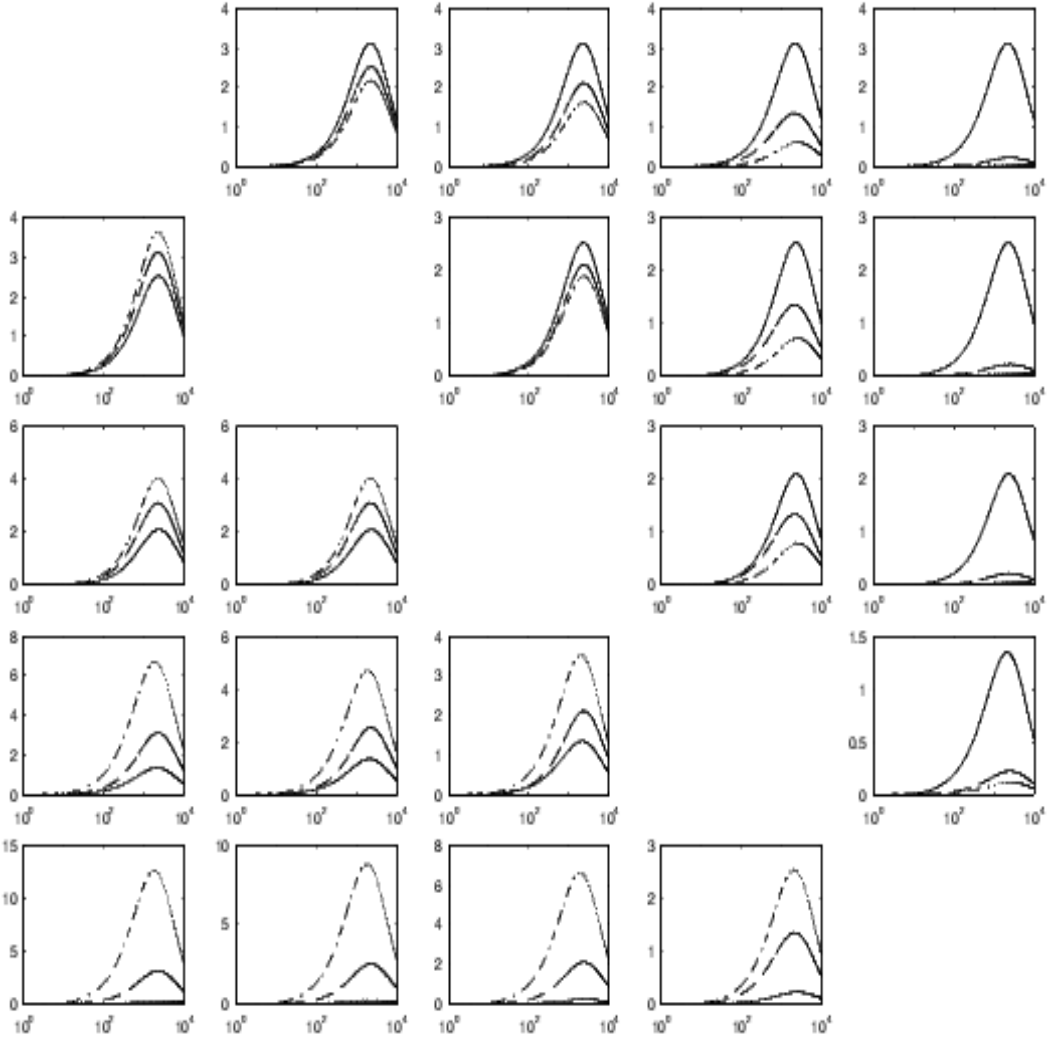, height=14cm, width=14cm, clip=}
\caption{The S-Z power spectrum calculated using unmatched TM
relations and $\sigma_8$ normalization. Units of the abscissa and 
ordinate are $\ell$ and $10^{-12}\,\ell(\ell+1)\,C_{\ell}/2\pi$, 
respectively. Rows 1-5 correspond to models 1,2,3,4,5. Columns 1-5
correspond to the correct $\sigma_8$ for the model. Solid curves
depict the `diagonal' terms, that is, calculations carried out with a 
specific TMR and its corresponding $\sigma_8$. Dash-dotted curves 
describe calculations made with the same TMR, but unmatched 
$\sigma_8$. Finally, the dashed curves illustrate the `corrected' 
power spectrum, i.e. keeping the unmatched $\sigma_8$ but now using 
the matching TMR.}
\label{fig:clm1}
\end{figure}

\begin{figure}
\centering
\epsfig{file=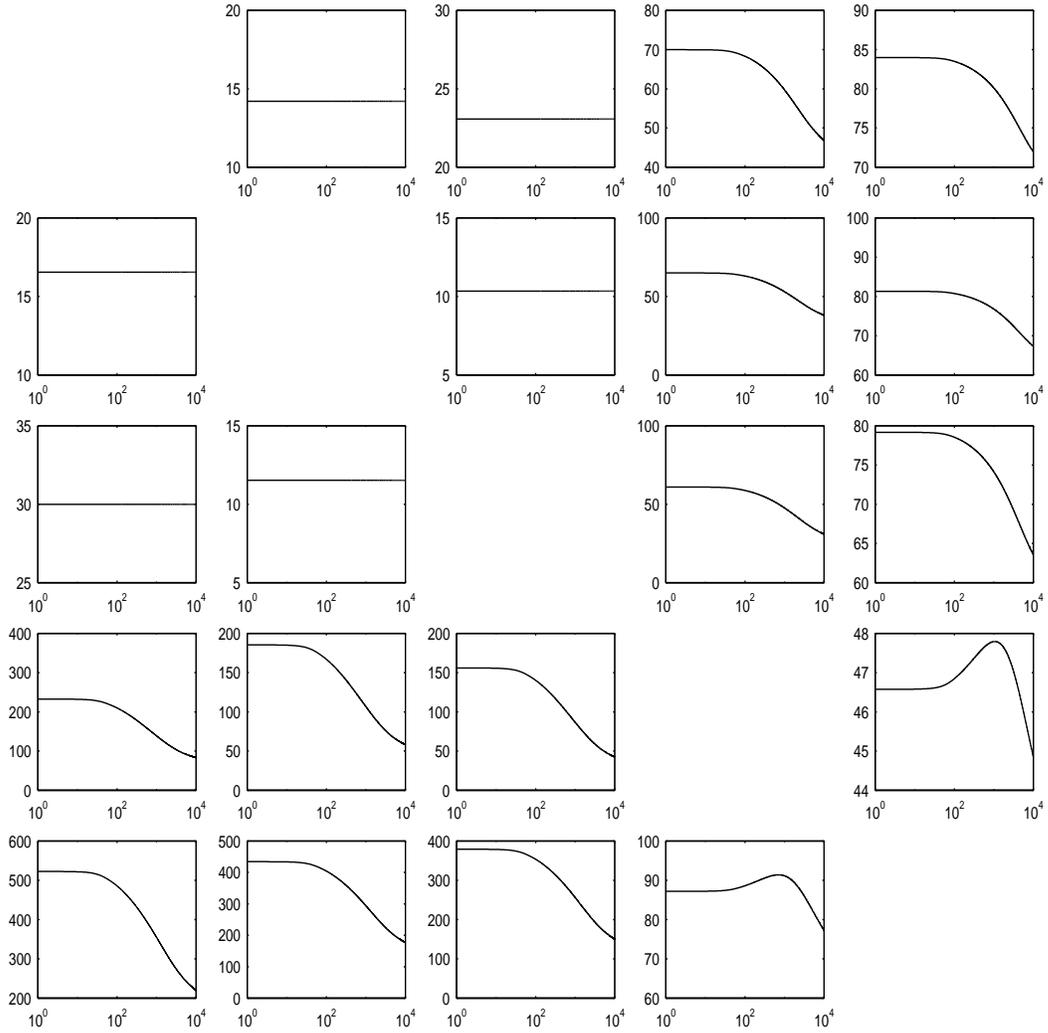, height=14cm, width=14cm, clip=}
\caption{As in figure (3), but here the relative differences 
between the dashed and dash-dotted curves (i.e., consistent
vs. inconsistent calculations) are specified in percentage units.}
\label{fig:clm2}
\end{figure}

The results of the second set of `cross' calculations are described in 
figure~\ref{fig:clm1}. Each panel contains three curves; the solid curves 
represent calculations employing a specific TMR and its corresponding 
$\sigma_8$. The dash-dotted curves show results of the calculations
employing the same TMR as for the solid curves, but a different $\sigma_8$.
Finally, the dashed curves show results for the same $\sigma_8$ as in the
dash-dotted curves, but with the TMR that corresponds to this parameter,
thereby providing a `correction' to the results illustrated by the
dash-dotted curves. In order to further clarify the results, the first
row will be considered as an example: In the first panel (2nd from the
left) the TMR of model 1 is used; here the temperatures are lowest, and
$\sigma_8$ assumes the highest value; the solid curve describes this case.
Now $\sigma_8$ is changed to the value extracted using model 2, i.e., a
lower value. However, in combination with the TMR of model 1, the power
spectrum is underestimated, since the TMR of model 2 should have been
used, rather than that of model 1. Indeed, model 2 predicts higher
temperatures than model 1, and therefore the correct power spectrum is
represented by the dashed curve. Obviously, as the difference between
the correct $\sigma_8$ and the one employed in the calculation increases,
the degree of underestimation (or overestimation) also increases, as can
be seen in the next panels of the first row. The relative errors
introduced by the inconsistent choice of these two elements can be readily
calculated; this is presented in figure~\ref{fig:clm2}. As anticipated,
the panels lying on both sides of the diagonal show lower relative errors
than those located further away from the diagonal, where differences
between adopted and correct values of $\sigma_8$ are larger. Errors of
up to a factor of $\sim 5$ are caused by this parameter mismatch. The
relative errors among the first three models are constant along the
multipole axis due to their functional uniformity. In other cases the
relative error changes with $\ell$.

The dependence of the S-Z power on $\sigma_8$ has already been discussed
in previous works, it is usually demonstrated by arbitrarily changing
its value. Although the change is not great (as the ones required
in order to induce a relative error of hundreds of percents as seen in
figure~\ref{fig:clm2}), a difference of $\lesssim 20\%$ (e.g.
$\sigma_8=0.9-1.1$) or more is common. Examples for large errors induced
by a change of $\sim 18\%$ in $\sigma_8$ may be seen in the third panels
of the third and fourth rows of figure~\ref{fig:clm2}: in the third row
the correct and unmatched values of $\sigma_8$ are $1.18$ and $1.0$,
respectively, and a relative error of $\sim 40-50\%$ appears around
$\ell=1000$. In the fourth row the correct and unmatched values are $1.0$
and $1.18$, respectively, and a relative error of $> 50\%$ is induced.

\begin{figure}
\centering
\epsfig{file=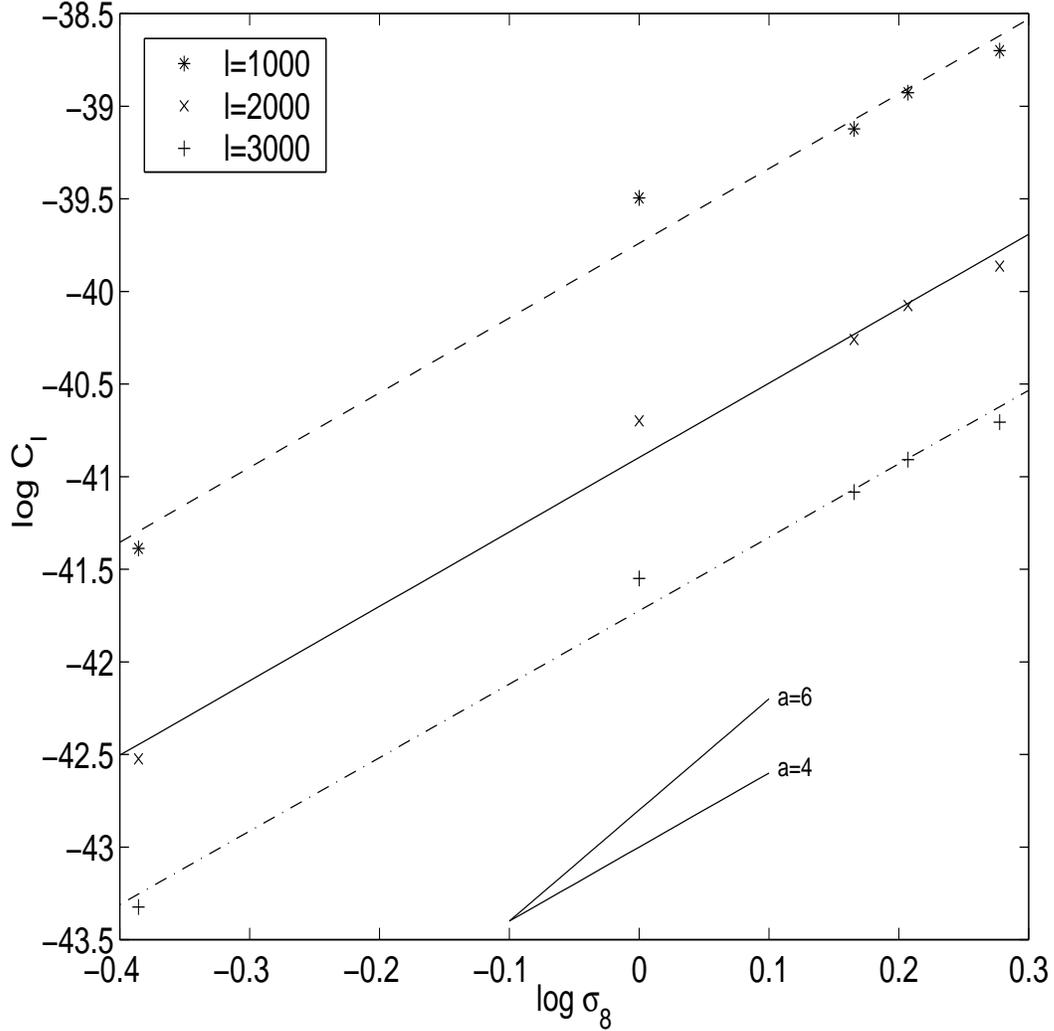, height=14cm, width=14cm, clip=}
\caption{The dependence of $C_{\ell}$ on $\sigma_8$, illustrated 
in a Log-Log plot. Asterices, x-marks and plus symbols represent 
$C_{\ell}$s at five different $\sigma_8$s, corresponding to models 
1-5, for $\ell=1000, 2000, 3000$. The dashed, solid, and dash-dotted
curves depict the best linear fits for each of the 5 data point sets. 
The two short lines appearing at the bottom of the plot are of slopes 
4 and 6, as indicated in the figure.}
\label{fig:clsig8}
\end{figure}

\begin{figure}
\centering
\epsfig{file=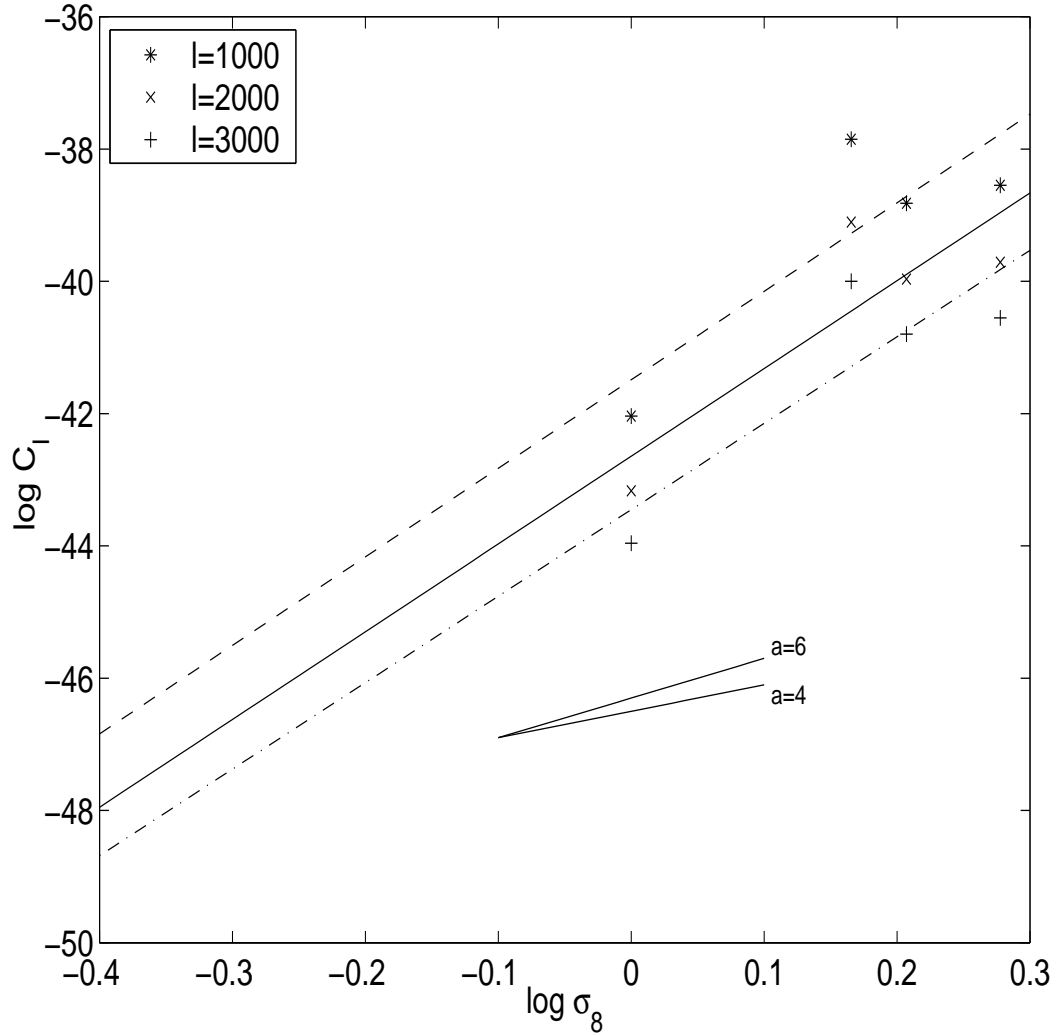, height=14cm, width=14cm, clip=}
\caption{The same as figure (5), but here the TMR models 
2-5 are used in conjunction with non-matching $\sigma_8$s, extracted 
from models 1-4, respectively. Thus, both higher values of $\sigma_8$ 
and temperatures scaled from the TMR are used, leading to an 
overestimation of the power spectrum.}
\label{fig:clsig82}
\end{figure}

The steep dependence of the S-Z power on $\sigma_8$ was quantified by
the proportionality $C_{\ell}\propto\sigma_{8}^{6-7}$ (e.g., Komatsu \&
Tetsu 1999). Our results indicate that this dependence is somewhat more
moderate, which is to be expected since an increase of $\sigma_8$ implies
a decrease in temperature by virtue of the anti-correlation between these
two parameters, as has been shown above and illustrated in
figure~(\ref{fig:t-sig}). For the three representative value of the
multipole (marking the range in which the peak of the S-Z power spectrum
occurs) $\ell=1000,\,2000,\,3000$, we plotted the corresponding $C_{\ell}$s
in models 1-5 against their corresponding $\sigma_8$. Best fits describing
the three sets of five data points all have slopes $\sim 4$, smaller than
the values ($\sim 6-7$) previously reported. This is illustrated in
figure~(\ref{fig:clsig8}). For comparison, we include
figure~(\ref{fig:clsig82}), in which the same analysis is applied, but
here the TMR models 2-5 are used in conjunction with the $\sigma_8$s
extracted using models 1-4. The corresponding power spectra are represented
by the dash-dotted curves in the four panels under the central diagonal of
figure~(\ref{fig:clm1}). Recall that the resulting power spectrum in this 
case is overestimated due to the fact that the TMR used in its evaluation 
would necessitate a lower $\sigma_8$ than the one used in the calculation. 
Thus, a steeper slope is expected in the $\log{C_{\ell}}-\log{\sigma_8}$ 
plot, which is indeed the case. The resulting slopes are actually steeper 
($\sim 13$) than $a\approx 6-7$, but the essential result of this analysis 
is that the scaling $C_{\ell}-\sigma_8$ is indeed sensitive to the adopted
TMR and its normalization.

\section{Dependence of Cluster Counts on $\sigma_8$ and the TMR}
\label{sect:szcc}

As is well known, the fact that the S-Z effect is independent of redshift
makes it an (almost) ideal tool for the detection of distant clusters.  
Cluster surveys are planned by several S-Z projects, culminating (according
to current plans) with the full sky mapping by the PLANCK satellite. The
143 GHz and 353 GHz channels of the HFI experiment on PLANCK are
particularly suitable for detecting large numbers of clusters. The S-Z
flux from a cluster is given by (e.g. Colafrancesco et al. 1997)
\begin{equation}
\Delta F_{\nu}(\gamma_{\ell})=\int
R_{s}(|\hat{\gamma}-\hat{\gamma_{\ell}}|,\sigma_{B})\,\Delta
I_{\nu}(\hat{\gamma})\,d\Omega, 
\end{equation}
where $\Delta I_{\nu}$ is the change of the CMB spectral intensity due 
to the thermal component of the S-Z effect. In the non-relativistic 
limit, $\Delta I_{\nu} = g(x) = x^{4}e^{x}(e^{x}-1)^{-2}
\left[x\coth(x/2)-4\right]$, where $T_0$ is the CMB temperature, and $y$ 
is the Comptonization parameter integrated along a los. The more 
accurate expression for $\Delta I_{\nu}$ includes relativistic corrections 
(Rephaeli 1995, Itoh \ea 1998, Shimon \& Rephaeli 2003). Since our main 
aim here is to display the dependence of cluster number counts on 
$\sigma_8$ and the TMR -- rather than precisely predicting the expected 
number of clusters -- we use the simpler non-relativistic limit of 
$\Delta I_{\nu}$. We can then write for the S-Z flux 
\begin{equation}
\Delta F_{\nu}=\D\frac{2(k_{b}T_0)^{3}}{(hc)^{2}}g(x)y_{0}
\int R_{s}(|\hat{\gamma}-\hat{\gamma_{\ell}}|,\sigma_{B})\cdot   
\zeta(|\hat{\gamma_{\ell}}|,M,z)\,d\Omega,
\end{equation}
where
\begin{equation}
\zeta(|\hat{\gamma_{\ell}}|,M,z)\equiv\D\frac{1}{\sqrt{1+(\theta/\theta_{c})^{2}}}\cdot\tan^{-1}
{\left[p\sqrt{\frac{1-(\theta/p\theta_{c})^{2}}{1+(\theta/\theta_{c})^{2}}}\right]}.
\end{equation}  
The measured signal is actually weighted over the spectral response of the
detector, $E_{\nu}$, usually approximated by a Gaussian, $G_{\nu}$. 
Therefore,
\begin{equation}
\Delta\overline{F}_{\nu}=\D\frac{\int \Delta F_{\nu}E(\nu)d\nu}{\int E(\nu)d\nu}.
\end{equation}
For simplicity, $G_{\nu}$ is taken here to be uniform over the passband 
centered on the central frequency. For the PLANCK/HFI $143\,$GHz channel, 
beam size is $8'$ and $\Delta\nu/\nu=0.25$. The predicted number of clusters 
with flux greater than $\Delta \overline{F}_{\nu}$ can now be calculated 
using
\begin{equation}
N(\Delta \overline{F}_{\nu})=\int r^{2}\frac{dr}{dz}dz \int_{\Delta \overline{F}_{\nu}}
B(M,z)N(M,z)dM.
\label{eq:counts}
\end{equation}  
The lower limit of the mass function integral is such as to correspond to 
the limiting flux from a cluster with mass $M$, which we take to be 
$30\,$mJy, the flux limit reachable in a one year observation. To 
evaluate this limiting mass, a binary function $B(M,z)$ is 
introduced. $B(M,z)=1$ if the flux corresponding to a given mass (and 
redshift) is greater than the limiting flux; otherwise, $B(M,z)=0$.
 
\begin{figure}
\centering
\epsfig{file=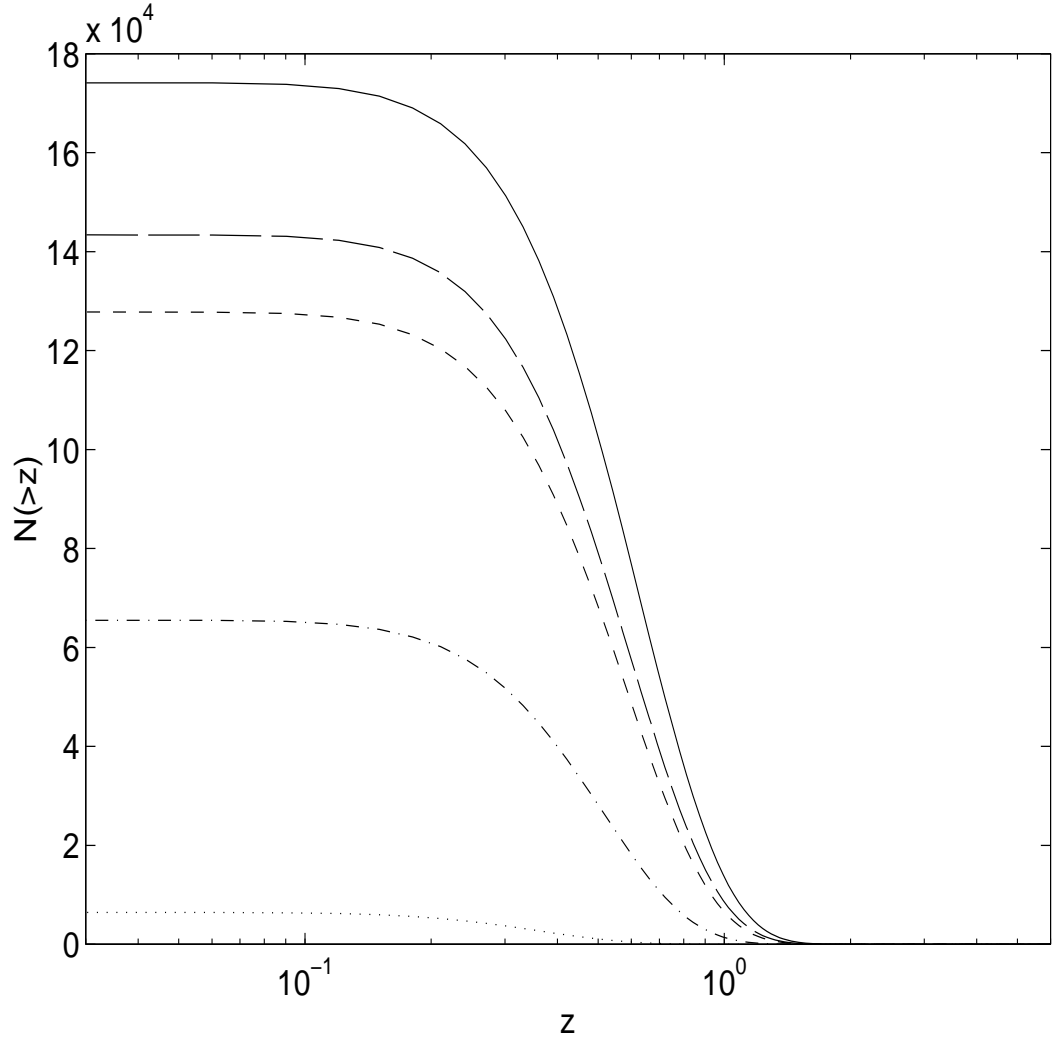, height=14cm, width=14cm, clip=}
\caption{S-Z cluster number counts calculated using the five TMR 
described above, and their matching values of $\sigma_8$. As in
figure (2), solid, long dashed, short dashed, dash-dotted 
and dotted curves correspond to models 1,2,3,4 and 5,
respectively.}
\label{fig:cnc}   
\end{figure}

\begin{figure}
\centering
\epsfig{file=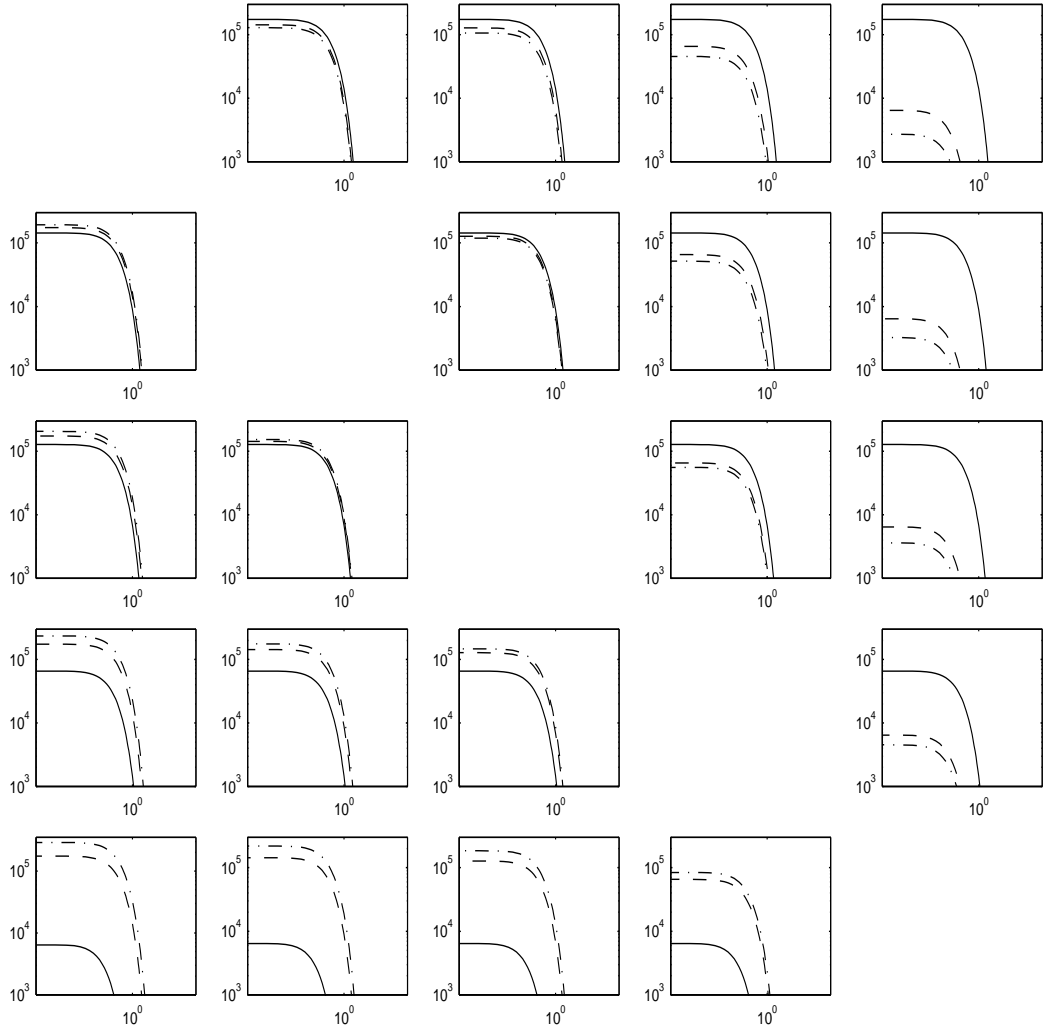, height=14cm, width=14cm, clip=}
\caption{Cluster number counts calculated using unmatching M-T relations and normalizations,
in likeness with figure (3). Abscissa and ordinate units are redshifts and 
counts.}
\label{fig:matrix3}
\end{figure}

The predicted number of detectable clusters with mass $M$ at redshift $z$ 
is proportional to the MF evaluated at these parameters. This depends on 
the integrated Comptonization parameter along all lines of sight to the 
cluster. Consequently, the arguments presented in \S\,2 are relevant also 
in this case. In figure~\ref{fig:cnc} the predicted cumulative number 
counts in models 1-5 are shown. A strong relation between the MF 
normalization and the cluster counts is clearly manifested. It may also 
be inferred from the figure that the largest contribution to the counts
comes from clusters lying at redshifts $\lesssim 0.4$. Since the
redshift distribution of clusters cannot be determined from S-Z measurements, 
the most relevant quantity is indeed the cumulative cluster counts at low
redshifts. Results of the `cross' calculations are presented in 
figure~\ref{fig:matrix3}. The curves in figure~\ref{fig:matrix3} resemble
those in figure~\ref{fig:clm1} in the sense that the differences between
the values represented by the `mismatched' and `correct' curves increase
farther away from the diagonal. The relative differences between the 
properly and improperly calculated numbers span a range of $\sim 10-70\%$ 
at the lowest redshifts, corresponding to the maximal cluster counts that 
would be detected in an all-sky S-Z survey. 

Although not observable, the larger differences exhibited at higher 
redshift reflect the higher sensitivity of the MF to $\sigma_8$ at the 
high-mass end. At such high redshifts only the most massive clusters would
be detected, owing to their larger fluxes, and therefore any change in the 
normalization that does not match the TMR used to evaluate the flux, would 
result in larger deviations of the counts with respect to a model employing 
a specific TMR and its corresponding normalization. A comparison between 
the results illustrated in figures~\ref{fig:clm1} and~\ref{fig:matrix3} 
suggests that an inconsistent choice of $\sigma_8$ and the TMR results in 
a rather pronounced over- or under-estimation of the power spectrum, and 
a milder (but significant) mis-estimation in the cumulative number counts, 
in particular at the lowest redshifts. This difference is clearly due to 
the steeper (gas) temperature dependence of the power ($\propto T_{e}^2$) 
as compared to the corresponding linear dependence of the flux.

\section{Conclusion}
\label{sect:conc}

Predicted profiles and levels of the angular power spectrum of the S-Z 
effect and of cluster number counts are strongly affected by modeling details 
and the particular choice of parameter values which explain the large 
variance in the values of these quantities in the literature. In this paper 
we focused on the tight dependence between the TMR and $\sigma_8$. We 
demonstrated that selecting these independently is generally not 
self-consistent and leads to considerable mis-estimation of the power 
spectrum and cluster counts. This can be is avoided when $\sigma_8$ is 
determined from CMB and large scale structure surveys that do not involve 
cluster TMRs.

Acknowledgement: This research has been supported by the Israeli
Science Foundation grant 729$\backslash$00-03 at Tel-Aviv University.

\newpage
\def\ref{\par\noindent\hangindent 20pt}

\newpage

\end{document}